\documentstyle[12pt]{article}
\begin{document}
\begin{flushright}
{CERN-TH.7411/94\\
NTUA.10/94\\
IOA.307/94}\\
\end{flushright}
\vspace*{3mm}
\begin{center}
{\bf  LOW ENERGY THRESHOLDS AND THE SCALAR MASS }\\
{\bf SPECTRUM IN MINIMAL SUPERSYMMETRY}\\
\vspace*{1cm}
{\bf G. K. Leontaris$^*$}\\
{\it CERN} \\
{\it Theory Division} \\
{\it 1211 Geneva 23} \\
 {\it Switzerland}\\
{\bf and}\\
\vspace*{0.3cm}
{\bf N.D. Tracas}\\
{\it Physics Department}\\
{\it National Technical University}\\
{\it GR-157 80 Zografou, Athens, Greece}\\
\vspace*{0.5cm}
{\bf ABSTRACT} \\
\end{center}
We discuss low energy threshold effects and calculate  the sparticle
masses in the context of the Minimal Supersymmetric Standard Model.
We pay particular attention to the top squark and the Higgs mass
parameters, and calculate the top Yukawa corrections, taking into
account the successive decoupling of each particle at its threshold.
We discuss the phenomenological implications in the context of the
radiative symmetry breaking scenario.
\noindent
\vspace*{2cm}
\noindent
\begin{flushleft}
{CERN-TH.7411/94\\
NTUA.10/94\\
IOA.307/94}
\\
August 1994
\end{flushleft}
\thispagestyle{empty}
\vfill
\hrule
$^*$Permanent address: Theoretical Physics Division, University
of Ioannina, \\GR 451 10 Ioannina, Greece.
\eject
\setcounter{page}{1}

The Minimal Supersymmetric Standard Model
\footnote{for reviews see for example \cite{susy}}
(MSSM) has by now been accepted as the most natural
extension of the Standard Model of Strong and Electroweak interactions.
As the recent experiments \cite{exp} approach closer to the energies
 where some of the
superpartners seems to acquire their masses,
 it is very important to obtain
higher  precision in the theoretical predictions of  the scalar
mass parameters, Yukawa coupling corrections, threshold effects etc.
In recent analyses \cite{LP1,GKL,LP2,BCPW,LT},
it has been shown that the semi-analytic procedure
in the calculations of the above quantities can offer the possibility of
investigating reliably the above effects. Moreover, the advantage
of analytic  expressions
 for the low energy parameters is more than obvious:
one can extract easily  information
about the role of the input parameters at the GUT scale ($m_0$, $m_{1/2}$,
$h_{t_G}$), since in the analytic procedure the low energy measurable
quantities can be expressed in terms of
calculable functions of the former with the boundary
 conditions incorporated
into these expressions.
Nevertheless, detailed theoretical predictions may be still pushed further
by estimating higher order effects, threshold corrections, etc.

In recent works it has been shown that the unification
 scenario\cite{DG} survives
even when various uncertainties arising from several sources (GUT and
low energy SUSY thresholds, input values for coupling constants,
experimental uncertainties etc.,)
are taken into account\cite{LP1,LP2,wa,wb}.
For example, in ref.\cite{LP1}, it has been shown that an effective
low energy  SUSY scale can be defined which, for a realistic mass
spectrum can  account for low energy  threshold effects.
More recently\cite{FG} a more accurate way to estimate the uncertainties
of such effects, which may also take into account two loop  corrections,
  has
been explored.

Predictions of low energy parameters turn out to be very
sensitive in all the above mentioned threshold effects.
In estimating  these effects, it has been
shown\cite{AKT} that
it is adequate to use the ``step'' approximation in the definition of
the beta function coefficients. Thus the values of the weak
mixing angle ($sin^2\theta_W$),
 the electromagnetic ($\alpha_{em}$) and the strong
 ($\alpha_3$) couplings, and other low energy parameters
can be given by their one loop formula with the addition
of two small correction
terms arising from two-loop and threshold effects in a semi-analytic
procedure\cite{LP1}.
 In calculating however the scalar masses themselves, one
should be  careful in particular for those affected
 from the top-Yukawa coupling
$h_t$. In this case due to the non-negligible contribution of
$h_t$, the evolution of
 $m_{\tilde t}^2$ squark and $m_{H_2}$ mass parameters
are determined by a coupled differential equation system.
Thus in addition
to the successive changes of the gauge coefficients as the sparticles
decouple, the boundary conditions
at each sparticle's threshold should be treated carefully.

 In the present analysis, we wish to investigate the above effects in the
context  of the MSSM
assuming that the gauge couplings unify in a simple  non-abelian gauge
group at an energy scale close to $10^{16}$GeV.
We find it useful to adopt a semi analytic procedure and provide
specific formulae for all the involved parameters, and compare them
with those of previous estimates where such effects were
not taken into account.

In particular the following issues will be discussed. We will start
assuming that the radiative symmetry breaking (RSB) scenario\cite{IR}
is an effective mechanism  operating in  the usual sense
at low energy, i.e. by driving one of the Higgs mass-squared parameters
negative at
energies close to $m_Z$. We are going to use one-loop corrections to the
effective
potential and estimate the effects in the $|\mu|$-parameter which plays an
essential role.
Next we calculate the exact contributions of
 the trilinear parameter $A$ and
compare the results with previous estimates
where these corrections were not included.
Finally, we are going to calculate the scalar masses for various choices
of the initial values $m_0,m_{1/2}$
 taking into account the afore-mentioned
threshold effects.

Starting at the GUT scale with a particular
 gauge group, one chooses specific
values for five independent parameters,
 namely $m_0$, $m_{1/2}$, $\mu$, $A$
and $B$. In the simplest case, all the scalars have a
universal \footnote{for recent discussions
where non-universal conditions at the GUT scale
are assumed see\cite{MN}}
mass $m_0$.
The masses evolve down to low energies
where one expects that one Higgs mass-squared parameter
becomes negative. This
triggers the $SU(2)\times U(1)$ symmetry breaking. The calculation of the
mass-squared parameters needed to check if this scenario is valid,
 requires
the solution of the coupled differential R.G. equations obeyed by these
parameters.

In the case of the small $\tan\beta\sim${\cal O}(1)
 scenario ($\tan\beta$ is
the ratio of the two v.e.v's),
 one may approximate the relevant differential
equations as follows
\begin{eqnarray}
\frac{dm^2_{\tilde t_L}}{dt}&=&\frac{1}{8\pi^2}
   \left(h^2_t\left(m^2_{H_2}+m^2_{\tilde t_L}
+m^2_{\tilde t_R}+A^2\right)-
      \sum^3_{i=1}c_i^Qg_i^2M_i^2\right)
\label{eq:demtl}\\
\frac{dm^2_{\tilde t_R}}{dt}&=&\frac{1}{8\pi^2}
   \left(2h^2_t\left(m^2_{H_2}+m^2_{\tilde t_L}
+m^2_{\tilde t_R}+A^2\right)-
      \sum^3_{i=1}c_i^Ug_i^2M_i^2\right)
\label{eq:demtr}\\
\frac{dm^2_{H_2}}{dt}&=&\frac{1}{8\pi^2}
   \left(3h^2_t\left(m^2_{H_2}+m^2_{\tilde t_L}
+m^2_{\tilde t_R}+A^2\right)-
      \sum^3_{i=1}c_i^Hg_i^2M_i^2\right)
\label{eq:demh2}\\
\frac{dm^2_{H_1}}{dt}&=&\frac{1}{8\pi^2}
                        \left(- \sum^3_{i=1}c_i^Hg_i^2M_i^2\right)
\label{eq:demh1}\\
\frac{dA}{dt}&=&\frac{1}{8\pi^2}
              \left(6h_t^2A-\sum^3_{i=1}c^A_ig_i^2M_i\right)
\label{eq:deA}
\end{eqnarray}
where only the top quark Yukawa coupling $h_t$ has been kept.
The coefficients $c_i$ are given by
\begin{eqnarray*}
c_i^Q=\{\frac{1}{15},3,\frac{16}{3}\}&\quad&\quad
c_i^U=\{\frac{16}{15},0,\frac{16}{3}\}\\
c_i^H=\{\frac{3}{5},3,0\}&\quad&\quad
c^A_i=\{\frac{13}{15},3,\frac{16}{3}\}
\end{eqnarray*}
while $M_i$'s are the gaugino masses and  $t=\ln Q$.

The differential equation for $H_1$ can be solved straightforwardly,
  since it is
independent from the others. The remaining four differential
equations define a coupled system which depends
strongly on the top Yukawa coupling. Making the identifications
\[
m_{\tilde t_L}\equiv \tilde m_1\quad{\mbox, }\quad\quad
m_{\tilde t_R}\equiv \tilde m_2\quad{\mbox {\rm and} }\quad\quad
m_{H_2}\equiv \tilde m_3
\]
the solution of the system is found to be
\begin{eqnarray}
m^2_{H_1}(t)&=&m^2_0+C_3(t)m_{1/2}^2
\label{eq:mh1}\\
\tilde m_n^2(t)&=&m_0^2+C_n(t)m_{1/2}^2-n\delta^2_m(t)-n\delta_A^2(t)
\label{eq:mn}\\
A(t)&=&q^{-1}(t)\left( A_G-m_{1/2}I_A(t)\right)
\label{eq:A}
\end{eqnarray}
where all the relevant functions are presented in the Appendix.

The scale dependence of the Yukawa coupling $h_t(t)$ can be found by
solving the RGE for that coupling
\begin{equation}
\frac{dh_t}{dt}=\frac{1}{8\pi^2}
\left(6h_t^2-\sum^3_{i=1}c_ig_i^2\right)h_t,
 \quad{\mbox {\rm where}}\quad
c_i=\{\frac{13}{15},3,\frac{16}{3}\}
\end{equation}
with the well known solution,
\begin{equation}
h_t(t)=h_{t_G}\gamma_U(t)q^{-1/2}(t)
\end{equation}
In all previous equations,
 the subscript $G$ denotes  the corresponding value
at the GUT scale.

There are four arbitrary parameters entering the above formulae, namely
$m_0$, $m_{1/2}$, $A_G$ and $h_{t_G}$.
 The first three of them, as was already
pointed out, are the soft input mass parameters at the unification scale
$M_G$.
 Since the above solutions enter the minimization of the Higgs potential
${\cal V}(H_1,H_2)$, their range can be phenomenologically constrained by
the requirement of generating a stable minimum for this potential.
 Of course,
a crucial role is played also by
 the fourth parameter, $h_{t_G}$, which should
be large enough to drive the $H_2$-Higgs mass-squared
parameter negative and
give a phenomenologically acceptable vacuum.

Experimental evidence\cite{CDF}
as well as theoretical expectations\cite{CK1}
 treating the Yukawa couplings
as dynamical variables, indicate
 that the top mass requires a large top-Yukawa
coupling,
 close to its infrared fixed point, i.e. $(m_t/\sin\beta)\sim 190$ GeV.
Therefore, since only the ratio $m_t/\sin\beta$ enters in the relevant
running scalar mass parameters, one may safely conclude that  in the fixed
point
solution for the top-mass,  their values depend mainly on the initial
values $m_0$ and $m_{1/2}$.
The above argument may be more transparent if one writes the above
mass-formulae
in the limit of the infrared fixed point of $h_t$.
 One then obtains\cite{LT}
\begin{equation}
\tilde m _n^2(t)=\left(1-\frac{n}{2}\right)m_0^2+
              \left[C_n(t)-\frac{n}{6}\frac{J(t)}{I(t)}\right]m^2_{1/2}
\label{eq:irf}
\end{equation}

The contribution of the trilinear parameter A introduces one more input
parameter at the GUT scale, but its role is less significant as long as
$A_G$ is of the order of $m_0$ (as expected). Indeed, by writing
$A_G=A_0m_0$, after some algebraic manipulations one can show that
$\delta^2_A$ is written as follows
\begin{equation}
\delta_A^2(t)=
\frac{1}{q(t)}\int_{t_G}^t\!q(t^\prime)d\left(\Delta_A^2\right)=
             -\frac{1}{q(t)}\int_{t_G}^t\!q(t^\prime)
                            \frac{h^2_t(t^\prime)}{8\pi^2}A^2(t^\prime)
                            dt^\prime
\end{equation}
Expanding $A^2(t)$, using Eq.(\ref{eq:A}), we write formally
\begin{equation}
\delta^2_A=\delta^2_{A_1}A_0^2m_0^2+
                      \delta^2_{A_2}A_0m_0m_{1/2}+
                                 \delta^2_{A_3}m^2_{1/2}
\end{equation}
where
\begin{eqnarray}
\delta^2_{A_1}(t)&=&
-\frac{1}{6}\frac{1}{q(t)}\left(\frac{1}{q(t)}-1\right)\\
\delta^2_{A_2}(t)&=&-\frac{1}{6}\frac{1}{q(t)}
                \left(\frac{1}{q(t)}I_A(t)-\gamma_A(t)\right)\\
\delta^2_{A_3}(t)&=&-\frac{1}{6}\frac{1}{q(t)}
              \left(\frac{1}{q(t)}I_A^2(t)-2\gamma_A(t)I_A(t)+
                        2I_A^\prime(t)\right)\\
I_A^\prime(t)&=&\int_{t_G}^t\!q(t^\prime)C_A(t^\prime)\gamma_A(t^\prime)
                            dt^\prime\\
\gamma_A(t)&=&\int_{t_G}^t\!C_A(t^\prime)dt^\prime=\sum_{i=1}^3
                                                       \frac{c_i^A}{b_i}
        \left(\frac{\alpha_i(t)}{\alpha_{i_G}}-1\right)
\end{eqnarray}

The coefficients $\delta_{A_{i}}$ depend on simple integrals of scale
dependent parameters.
 Evaluation of the relevant integrals give the following results
(for $m_t\sim 175$ GeV and SUSY breaking $M_S=(500-1500)$ GeV)
\begin{equation}
\delta^2_A=0.005A_0^2m_0^2-
                      0.020A_0m_0m_{1/2}+
                                 (0.146-0.155)m^2_{1/2}
\end{equation}
which add too small corrections to the solutions where $A$ was ignored.
Moreover, these corrections become even smaller\cite{GKL,BCPW}
as  $m_t$ approaches its infrared fixed point value.

In the above equation, the simplified assumption
was made that all scalar masses decouple at the same scale, namely $M_S$.
In the running of the RGE's it is assumed that there is a great ``desert''
between the GUT and the weak scale while a low energy SUSY scale $M_S$
is assumed so that for energies lower than $M_S$ one uses the
Standard Model beta function- and $c_i$ coefficients.
However, as has been already pointed out,
a more careful treatment should also take into account threshold effects
due the successive decoupling of these scalar masses from the spectrum
at different scales.
In the semi-analytic approach of ref.\cite{LP1} an effective scale
$M_S^{eff}$ was assumed which can be roughly estimated to be
\begin{equation}
M_S^{eff}=
\left(\frac{\alpha_2(m_{\tilde W})}
{a_3(m_{\tilde g})}\right)^{\frac{28}{19}}|\mu|\approx \frac{1}{5}|\mu |
\end{equation}
to account for the SUSY scalar mass effects.
The $|\mu |$ parameter can also be given in terms of known parameters by
solving the minimization conditions of the neutral Higgs potential.
Taking also into account one-loop contributions to the superpotential,
we may obtain the following approximate formula for $|\mu|$
{\cite{AN92}}
\begin{equation}
|\mu|=\sqrt{(\mu _0^2+\eta ^2)/(1-\Omega ^2)}
\label{eq:newmu}
\end{equation}
In deriving the above, the approximation
$\ln m^2_{\tilde t_1}\sim \ln m^2_{\tilde t_2}\sim
\ln <m^2_{\tilde t}>$
has been used ($m^2_{\tilde t_1}$ and $m^2_{\tilde t_2}$ are the
eigenvalues of the stop mass matrix).
$|\mu_0|$ is the tree level contribution while $\eta$ and $\Omega$
are defined in the Appendix.
For
$tan\beta \ge 1.1$, $|\mu|$ is less than $1.5$ GeV.  One therefore,
for sensible  values  of the parameter
 $|\mu |\le (1-2)$ TeV, could define a reasonable  scale $M_S^{eff}$; below
that
scale the beta function
coefficients should turn to their non-supersymmetric
form.

The precise effects, however, are found by the successive change of all the
beta function dependent coefficients at each particle's threshold.
Assuming only one-loop corrections, in the Minimal Supersymmetry with
three families and two Higgses, one can write the $b_i's$ in the following
form\cite{LP1,FLY,HRS}
\begin{eqnarray}
b_1&=&
\frac{4}{3}n_g+\frac{1}{10}n_H^{SM}+\frac{2}{5}\theta_{\tilde{H}}
+\frac{1}{10}\theta_{{H_2}}
\nonumber \\
&+&
\frac{1}{5}\sum _{i=1}^3\left[\frac{1}{12}
(\theta_{\tilde{u}_{L_i}}+\theta_{\tilde{d}_{L_i}})
 +\frac{4}{3}\theta_{\tilde{u}_{R_i}}+
\frac{1}{3}\theta_{\tilde{d}_{R_i}}+
\frac{1}{4}(\theta_{\tilde{e}_{L_i}}+
\theta_{\tilde{\nu}_{L_i}})+\theta_{\tilde{e}_{R_i}}\right]\\
b_2&=&-\frac{22}{3}+\frac{4}{3}n_g+\frac{1}{6}n_H^{SM}+
\frac{4}{3}\theta_{\tilde{W}} \nonumber \\
&+&\frac{2}{3}\theta_{\tilde{H}}
+\frac{1}{6}\theta_{{H_2}}+\frac{1}{2}\sum _{i=1}^3
(\theta_{\tilde{u}_{L_i}}\theta_{\tilde{d}_{L_i}}
+\frac{1}{3}\theta_{\tilde{e}_{L_i}}
\theta_{\tilde{\nu}_{L_i}})\\
b_3&=&{-11}+\frac{4}{3}n_g\nonumber \\
&+&2\theta_{\tilde{g}}+
\frac{1}{6}\sum _i^3\left[
\theta_{\tilde{u}_{L_i}}+
\theta_{\tilde{u}_{R_i}}+
\theta_{\tilde{d}_{L_i}}+
\theta_{\tilde{d}_{R_i}}
\right]
\end{eqnarray}

In the above formulae for $b_i$'s, $\tilde H$ stands for the
higgsino contribution $\tilde W$ for the winos, etc., while
for any particle's threshold with mass $m^2_{s_i}$, we have
denoted  $\theta_{s_i}\equiv \theta (Q^2-m^2_{s_i})$.

In our semi-analytic approach,
when evolving the gauge and Yukawa couplings
as well as the scalar mass parameters  down to low energies,
we find it sufficient to define the following $b_i$-changing scales:
We assume a common scale $Q_L$ for the decoupling of $\tilde u_{L_{1,2}},
\tilde d_{L_{1,2}}$ sparticles, while we assume that they are not
very different from their mass eigenstates. The next scale is
the one defined as an average scale $Q_R$ of their right-handed partners.
In the case of the universal scalar masses at the GUT scale, however,
these two scales could not differ substantially unless $m_{1/2}$
is very large ($\ge$ 1TeV). Therefore, the thresholds arising between these
two scales are not expected to have a significant effect.

We define as a third scale (subsequently denoted with $t_1=ln{\tilde m_1}$)
where we  change the $b_i$ and $c_i$ coefficients, the scale
where the $t_L$-squark acquires its mass.
 As it is expected,
due to the large negative contributions from the top-quark Yukawa coupling,
this mass should
be substantially smaller than those of the $L,R$ squarks of the
first two generations. Finally, we define two more new scales above
the weak and top mass scale,
 namely the
 scale were the $t_R$ squark gets
 its mass (subsequently denoted with $t_2=ln{\tilde m_2}$) and
an average scale for all other contributions (sleptons etc).
The hierarchy of these two latter
 scales depends strongly on the point
 of the parameters space $(m_0,m_{1/2})$
 one has chosen. A simple inspection of the obtained
 evolution equations for their masses shows that
 slepton masses are larger than
$m_{\tilde t_R}$ for  $m_0$ values much bigger
 than $m_{1/2}$ while the opposite is true when $m_{1/2}> m_0$.
Of course all the above scales
should be carefully incorporated in the analytic
formulae presented above.
 All the relevant integrals (see Appendix) should split
into sums over the various scales.

Our next step is the determination
 of the $m_{\tilde t_L}$ and $m_{\tilde t_R}$
mass parameters which also define the afore-mentioned scales
 where the beta function coefficients should also change.
For a given ($m_0,m_{1/2}$) pair, one can compute the left
 and right squark masses of the first two generations.
The negative corrections $\delta_m^2(t_1)$ may then determine
  the mass of the $\tilde t_L$ squark. Below this scale,
 top-Yukawa negative corrections should not include contributions
 from diagrams involving  $\tilde t_L$.
Therefore in the range defined by
$m_{\tilde t_L}\ge Q \ge m_{\tilde t_R}$
(i.e.  for the range $(t_1,t_2)$) the evolution equations for
 $m^2_{\tilde t_R}\equiv \tilde m_2^2$
and $m^2_{H_2}\equiv \tilde m_3^2$
can be written as follows
\begin{eqnarray}
{\tilde m}^2_{2}&=&
   m^2_{2}(t_1)+ C_2(t_1,t)\tilde m^2_{1/2}-2\delta (t_1,t)
\label{eq:delm2}\\
{\tilde m}^2_{3}&=&
   m^2_{3}(t_1)+ C_3(t_1,t)\tilde m^2_{1/2}-3\delta (t_1,t)
\label{eq:delm3}
\end{eqnarray}
where $\tilde m^2_{n}(t_1)$ are the mass parameters calculated at the
scale $t_1=ln(m_{\tilde t_L})$. Top-Yukawa corrections $\delta (t_1,t)$
contain now  the sum only of $t_R$ squark and the $H_2$ Higgs mass
parameters, and possibly (depending on the specific values
 of the  ($m_0,m_{1/2}$) pair)
 the $A(t)$ trilinear parameter,
\begin{eqnarray}
 \delta (t_1,t)= \int^t_{t_1} \frac{h^2_t(t^\prime)}{8\pi^2}
\left(\sum^3_{n=2}m^2_{n}(t^{\prime})+
\theta (t^{\prime}-t_A)A^2(t^{\prime})\right)dt^{\prime}
\end{eqnarray}
where $t_A$ defines the logarithm of the scale at which
 the trilinear mass parameter stops running.
The above corrections can
be calculated easily, by solving (\ref{eq:delm2}),
(\ref{eq:delm3}). The result is
 \begin{eqnarray}
\delta (t_1,t)= -q^{-\frac{5}{6}}(t_1,t)
\int^t_{t_1}q^{\frac{5}{6}}(t_1,t^{\prime})u_0(t^{\prime})
\frac{h_t^2(t^{\prime})}{8\pi^2} dt^{\prime}
\end{eqnarray}
where
\begin{eqnarray}
u_0(t)=2m_0^2+\left[C_2(t_1,t)+C_3(t_1,t)\right]m_{1/2}^2-5
[\delta_m^2(t_1)-\delta_A^2(t_1)]
\end{eqnarray}
After the  decoupling from the massless spectrum of the $t_R$ squark
at the scale $t_2=ln(m_{\tilde t_R})$,
 one ends up with only the Higgs mass
parameter whose evolution from $t_2$ until the
 transmutation scale is given by
the formula
 \begin{eqnarray}
\tilde m_3^2(t)=\tilde m_3^2(t_2)+C_3(t_2,t)m^2_{1/2}-3\,
 q^{-\frac{1}{2}}(t_2,t)
\int^t_{t_2}q^{\frac{1}{2}}(t_2,t^{\prime})v_0(t^{\prime})
\frac{h_t^2(t^{\prime} )}{8\pi^2}
dt^{\prime}
\end{eqnarray}
where
\begin{eqnarray}
v_0(t)=\tilde m_0^2(t_2)+C_3(t_2,t)m_{1/2}^2
\end{eqnarray}

In tables $I$ and $II$ we present the calculated scalar mass spectrum
for two initial values of the $h_{t_G}$, (case $II$, very close to its
 infrared fixed point value), and
representative choices of $m_0,m_{1/2}$ pairs.
The last column of these tables shows the prediction for
$\alpha_s$ low energy parameter, for each sparticle spectrum.
 In obtaining our results,
 we have worked in the low $tan\beta$ regime while
we have allowed $~10\%$ deviations from the GUT relation
$h_b(t_G)=h_{\tau}(t_G)$.
On the other hand, the obtained values for $sin^2\theta_W$ and $m_t$ are
consistent with the relation
$sin^2\theta_W(m_Z)=0.2324-10^{-7}\times
\left\{ \left(m_t/{\mbox{\rm GeV}}\right)^2-143^2   \right\}\pm 0.0003$.

As it can be inferred from the tables, the $t_{L,R}$ squarks and the
average sparticle spectrum is lighter for larger top couplings and
 top mass.
 When $m_0$ is relatively small,  slepton masses are the
lighter sparticles while $m_{\tilde t_R}$ becomes  light when
$m_{1/2}\ll m_0$. Of cource, right sleptons (not shown in the tables)
have slightly smaller masses than their left partners.


\begin{center}
\vglue 0.4cm
\begin{tabular}{||cc|ccccc|c||}    \hline \hline

$m_0$ &$m_{1/2}$ &$m_{Q_L}$ &$m_{Q_R}$&$m_{\tilde t_L}$ &$m_{\tilde t_R}$
&$m_{\tilde l}$&$\alpha_s$ \\ \hline \hline

338 &$  423$  &$950  $ &$ 915$&853&697&450&.118\\ \hline
250 &$  359$  &$800  $ &$ 770$&722&596&357&.117\\ \hline
255 &$  306$  &$700  $ &$ 675$&627&513&335&.116\\  \hline
110 &$  324$  &$700  $ &$ 672$&640&539&256&.115\\ \hline
182 &$  189$  &$450  $ &$ 435$&400&325&225&.114\\  \hline
322 &$  100$  &$400  $ &$ 395$&316&192&340&.114\\ \hline
311 &$  100$  &$380  $ &$ 375$&301&187&319&.114\\ \hline
280 &$   95$  &$350  $ &$ 345$&280&178&288&.114\\  \hline
246 &$  113$  &$350  $ &$ 343$&290&204&340&.114\\ \hline
\hline
\end{tabular}
\vglue 0.2cm
{\bf Table I.}
{\it
The supersymmetric sparticle spectrum together with the corresponding
$\alpha_s$ prediction
in the low $tan\beta$ scenario
($tan\beta \sim 1.5$) and $m_t\le 165$ GeV, fixing $sin^2\theta_W$
around its central value $\sim .232$}
\end{center}
Now let us discuss the effects of the decoupling of the havier quarks
in the rest of the sparticle spectrum.
The successive decoupling of each scalar mass term from the relevant
differential equation, has modified the negative corrections
induced by the top-coupling below the $t_L$ squark mass.
In the low $tan\beta $ scenario this treatment has a direct effect
only on the $t_R$-squark and the $m_{H_2}^2$ mass parameters.
In particular, in
our semi-analytic treatment we observe that the $\tilde t_R$ mass
has increased by
$(1-5)\%$ , relative to a naive treatment of the boundary
conditions at each particle's threshold\cite{GKL},
depending on the specific choice of the $(m_0, m_{1/2})$ point.
Such corrections are therefore of the same order but with the
opposite sign of the $A$-trilinear term corrections given
by the formula (\ref{eq:A}).
The maximal change occurs when  the top Yukawa coupling
gets closer to its fixed point value ({\it Table II}).
The effect however is still small, since the modified equation
 (\ref{eq:delm2}) runs only on a  small
region, namely between the $\tilde m_{t_L}$ and $\tilde m_{t_R}$ scales.
A larger effect is found in the $m_{H_2}$ running mass whenever
the transmutation scale is  substantially  different from
 $\tilde m_{t_L}$.

Finally, as it has been pointed out above, the $\tilde t_R$ mass is
smaller than the slepton masses in specific regions of the
($m_0,m_{1/2}$) parameter space.
A qualitative picture of the $m_{\tilde t_{R}}$ and $m_{\tilde l}$
variation in terms of $m_0$ and $m_{1/2}$ is given in the Figure.


\begin{center}
\vglue 0.4cm
\begin{tabular}{||cc|ccccc|c||}    \hline \hline
$m_0$ &$m_{1/2}$ &$m_{Q_L}$ &$m_{Q_R}$&$m_{\tilde t_L}$ &$m_{\tilde t_R}$
 &$m_{\tilde l}$&$\alpha_s$ \\ \hline \hline
251 &$  359$  &$800  $ &$ 770$&718&588&356&.117\\ \hline
256 &$  306$  &$700  $ &$ 675$&624&505&335&.116\\  \hline
488 &$  236$  &$700  $ &$ 685$&576&395&515&.118\\ \hline
638 &$  135$  &$700  $ &$ 695$&525&241&645&.119\\ \hline
 73 &$  277$  &$600  $ &$ 577$&547&451&450&.114\\ \hline
 53 &$  254$  &$551  $ &$ 529$&503&424&185&.114\\  \hline
182 &$  188$  &$450  $ &$ 435$&399&320&226&.116\\ \hline
322 &$  102$  &$400  $ &$ 395$&313&180&340&.114\\ \hline
\hline
\end{tabular}
\vglue 0.2cm
{\bf Table II.} {\it
The supersymmetric sparticle spectrum together with the corresponding
$\alpha_s$ prediction for  $h_{t_G}$ Yukawa
coupling close to its non-perturbative value, and $m_t \sim 170$ GeV.
Again $sin^2\theta_W$ has been fixed to its central value
$\sim .232$}
\end{center}

We can see that there is a
considerable fraction of the parameter space $(m_0,m_{1/2})$
 which allows solutions
of relatively  small
$m_{\tilde t_R}$. For  example, in the last entry of table II,
 $m_{\tilde t_R}$
is of the order of the top-quark mass. This would imply that,
after the diagonalisation of the squark mass matrix, the light
physical mass eigenstate $m_{\tilde t_1}$
 could be  as small as $150$ GeV.
 This gives hope that future experiments may discover supersymmetric
signatures.

To summarize our results, we have used a semi-analytic approach
to calculate the supersymmetric spectrum in the small $\tan\beta$
regime, taking into account low energy threshold effects.
We have given special emphasis to top squark and Higgs mass
parameter calculation, which in the presence of a heavy top quark
receive large negative contributions. We have examined in detail
the effects of the `decoupling' of the heavier sparticles from the
renormalization group equations of the lighter ones,
 and we have found that
our treatment of the boundary conditions, results in an increase
$(1-5)\%$ of the $\tilde t_R$
mass parameter, compared to a naive treatment.
Corrections on the scalar masses from the trilinear
parameter $A$ are treated also analyticaly and
found to be of the same order for moderate initial values
($A_G\sim -\sqrt{3}m_0$)
Furthermore, we have examined general properties of the sparticle spectrum
and observed interesting correlations.
Thus, large values of $m_0$ compared
to that of $m_{1/2}$ imply that $m_{\tilde t_R}$ is lighter than the
left slepton masses, while the opposite is true for $m_{1/2}>m_{0}$.
Moreover, in the small $tan\beta$ regime that we are examining here,
for a considerable
fraction of the $(m_0,m_{1/2})$ space,
a light t-squark ($\sim 150$ GeV) can be obtained,
 which might be found in
accessible energies by experiment in the near future.

\vspace{1cm}
We wish to thank C. Kounnas for helpful discussions. The work of N.D.T
is partially supported by a C.E.C. Science program SC1-CT91-0729.

\newpage

{\bf APPENDIX}

The scale dependent coefficients in the scalar mass solutions
(\ref{eq:mh1},\ref{eq:mn},\ref{eq:A}) are given
by the following general formula
\begin{equation}
C_n(t_1,t)=\sum^3_{i=1}
\frac{c_i^n}{2b_i\alpha^2_{i_G}}
\left(\alpha_i^2(t)-\alpha_i^2(t_1)\right)
,\quad\quad n=1,2,3
\end{equation}
with the identifications $C_1\equiv C_Q$, $C_2\equiv C_D$
and $C_3\equiv C_{H_2}$. We define $C_n(t)\equiv C_n(t_G,t)$.

The gauge dependent functions $\gamma_U(t)$, and $q(t)$
which arise from the top-Yukawa differential equation have the form
\begin{equation}
\gamma_U(t)=\prod_{n}
\prod_{i=1}^3\left(\frac{\alpha_{i}(t_n)}{\alpha_i(t_{n-1})}\right)
^{c_i^n/2b_i^n}
\label{eq:gammaU1}
\end{equation}
\begin{equation}
q(t_1,t)=1+\frac{3h^2_{t_G}}{4\pi ^2}\: I(t)=
     1-\frac{3h^2_{t_G}}{4\pi ^2}
            \int^{t}_{t_1}\gamma_U^2(t^\prime)dt^\prime
\label{eq:q}
\end{equation}
where the index $n$ runs over all
 the intermediate scales. Again we define
$q(t)\equiv q(t_G,t)$.

The negative Yukawa contributions of Eq.(\ref{eq:mn}) are found to be
\cite{GKL}
\begin{equation}
\delta_m^2(t)=\left(\frac{m_{top}(t)}{2\pi v\gamma_U\sin\beta}\right)^2
               \left(3m^2_0I(t)+m^2_{1/2}J(t)\right),
\label{eq:deltam}
\end{equation}
\begin{equation}
\delta_A^2(t)=\Delta^2_A(t)-\frac{3}{2}
          \left(\frac{m_{top}(t)}{\pi v\gamma_U\sin\beta}\right)^2E^2_A
\label{eq:deltaA}
\end{equation}
where $ v=246$ GeV, while the quantities $I,J,I_A$
are functions of scale dependent integrals given by:

\begin{equation}
I(t)=-\int^{t}_{t_G}\gamma_U^2(t^\prime)dt^\prime,
\quad\quad
J(t)=
   -\sum^3_{i=1}\int^t_{t_G}\!\gamma_U^2(t^\prime) C_i(t^\prime)
                                                            dt^\prime
\label{eq:IJ}
\end{equation}
\begin{equation}
\Delta_A^2(t)=\int_t^{t_G}\!\frac{h^2_t(t^\prime)}{8\pi^2}A^2(t^\prime)
                dt^\prime,
\quad\quad
E_A^2(t)=\int_t^{t_G}\!\gamma_U^2(t^\prime)\Delta_A^2(t^\prime)dt^\prime
\label{eq:DeltaAEA}
\end{equation}
\begin{equation}
I_A(t)=\int^t_{t_G}\!q(t^\prime)C_A(t^\prime)dt^\prime
      =\frac{1}{2\pi}\sum_{i=1}^3c^A_i\alpha_{i_G}
      \int^t_{t_G}\!q(t^\prime)\frac{\alpha_i^2(t^\prime)}{\alpha^2_{i_G}}
                                                  dt^\prime
\label{eq:IA}
\end{equation}

The minimization conditions of the tree-level neutral Higgs potential
give the following solution for the $|\mu_0|$- parameter\cite{LT}
\begin{eqnarray}
|\mu_0 | = \frac{1}{\sqrt{2}} \Big \{\frac{k^2+2}{k^2-1} m_0^2
+ \Big ( \frac{k^2}{k^2-1} \frac{J}{I} - 1\Big ) m^2_{1/2}
 - M^2_Z\Big \}^{1/2}
\label{eq:mu}
\end{eqnarray}
with $k = \tan\beta$.

The parameters $\eta ,\Omega$ entering the one loop formula are given by

\begin{eqnarray}
\eta ^2&=&\frac{\alpha_2}{8\pi\cos^2\theta_W}
\left\{\left[\left(\frac{1}{4}-\rho^2\right)
\left(M^2_{LL}+M^2_{RR}\right) \right.\right.
 \nonumber\\
&+& \left.
\left(M^2_{LL}-M^2_{RR}\right)
          \left(\frac{1}{4}-\frac{2}{3}\sin^2\theta_W\right)
-\rho^2A^2\right]\left(\ln\tilde \rho^2-1\right)\nonumber\\
&-&\left.
2m^2_t\left(\ln\rho^2-1\right)\frac{\rho^2}{k^2}\right\}
\frac{k^2+1}{k^2-1}\\
\Omega^2&=&\frac{\alpha_2}{8\pi\cos^2\theta_W}
\left\{\frac{\rho^2(k^2+1)}{k^2(k^2-1)}\right\}
\left(\ln\tilde \rho^2-1\right)
\label{eq:eta}
\end{eqnarray}
with $\rho=m_t/M_Z$, $\tilde \rho=<m_{\tilde t}>/M_Z$ and $\mu_0$ the
tree level parameter defined in (\ref{eq:mu}).
Finally the $t$-squark mass-combinations $M^2_{LL}\pm M^2_{RR}$
are given by

\begin{eqnarray*}
M_{LL}^2+M_{RR}^2&=&\frac{1}{2}m_0^2
                    +(C_1+C_2-\frac{J}{2I})m_{1/2}^2+2m_t^2+
                      \frac{1}{2}m^2_Zcos 2\beta\\
M_{LL}^2-M_{RR}^2&=&\frac{1}{2}m_0^2
                     +(C_1-C_2+\frac{J}{6I})m_{1/2}^2+
                       (\frac{4}{3}M^2_W-\frac{5}{6}M^2_Z)cos 2\beta
\end{eqnarray*}


\noindent
{\bf Figure Captions}

\vspace{1cm}

{\bf Figure}. Surfaces of  $m_{\tilde l}$ and  $m_{\tilde t_R}$
showing their variation as a function of $m_0$ and $m_{1/2}$.
\end{document}